\documentclass[12pt]{spieman}  
\usepackage{amsmath,amsfonts,amssymb}
\usepackage{cleveref}
\usepackage{graphicx}
\usepackage{subfigure}
\usepackage{makecell}
\usepackage{booktabs}
\usepackage{multirow}
\usepackage{tikz}
\usetikzlibrary{decorations.pathreplacing}
\usepackage{setspace}
\usepackage{orcidlink}

\usepackage{upgreek}

\usepackage{lineno}
\usepackage{soul}

\newcommand{\um}{$\upmu$m}

\title{Bright Region Reset: an on-detector strategy for minimizing the impacts of atmospheric emission lines on spectral observations}

\author[a,*]{Theodore A. Grosson\orcidlink{0000-0002-5729-8894}}
\author[b]{Edward L. Chapin\orcidlink{0009-0003-8033-1351}}
\author[b]{Tim Hardy\orcidlink{0009-0006-0855-8422}}
\author[c,a]{Masen Lamb\orcidlink{0000-0002-2621-9655}}
\author[b]{Jordan Lothrop}
\author[b,a]{Alan W. McConnachie\orcidlink{0000-0003-4666-6564}}
\author[b]{Richard Murowinski}
\affil[a]{University of Victoria, Department of Physics \& Astronomy, Victoria, V8P 5C2, Canada}
\affil[b]{NRC Herzberg Astronomy and Astrophysics Research Centre, Victoria, V9E 2E7, Canada}
\affil[c]{NSF's NOIRlab / International Gemini Observatory, Hilo, 96720, USA}

\graphicspath{{img/}}

\begin{document} 
\maketitle

\begin{abstract}
Observations in the near-infrared using large ground-based telescopes are adversely impacted by bright atmospheric emission lines, particularly the OH Meinel bands. These lines can saturate a moderate-resolution spectrograph on the order of minutes, resulting in information loss at the wavelengths of the lines. OH lines also vary on similar timescales, requiring frequent sky exposures to be able to subtract the sky spectrum from that of the target. In this paper we present a new method, which we call bright region reset (BRR), to prevent the saturation of these lines in near-infrared spectra while simultaneously improving information about their variability. This is accomplished by periodically resetting pixels that contain bright lines on a detector capable of sub-window readout while the rest of the detector continues integrating. This method is demonstrated on the McKellar Spectrograph in the 1.2~m telescope at the Dominion Astrophysical Observatory in Victoria, Canada. Using a Teledyne H2RG detector, we reset the emission lines produced by an arc lamp while still recording their flux. We show no degradation in the resulting spectrum compared to a conventional observing mode. Unlike other OH line mitigation strategies, the BRR method not only avoids loss of information at wavelengths containing the lines, but also provides higher-cadence information on sky line variability, making it a promising technique for implementation at observatories. We advocate demonstrating this method on sky at existing 8--10~m class facilities with near-infrared spectrographs equipped with HxRG detectors in order to test its feasibility for use in sky subtraction schemes for premier modern spectrographs, including the upcoming generation of instruments for the Extremely Large Telescopes.
\end{abstract}


{\noindent \footnotesize\textbf{*}Theodore A. Grosson,  \href{mailto:tgrosson@uvic.ca}{tgrosson@uvic.ca} }


\section{Introduction}
\label{sec:intro}

Near-infrared (NIR) spectroscopy is an invaluable asset for modern astronomy. For example, the James Webb Space Telescope (JWST) Near-Infrared Spectrograph (NIRSpec) and Very Large Telescope (VLT) Enhanced Resolution Imager and Spectrograph (ERIS) are providing breakthrough observations from the high-redshift universe to nearby systems in the Milky Way\cite{Boker2022,Birkmann2022,davies_enhanced_2023}. Several new, ground-based NIR spectrographs are in various stages of development, including the VLT Multi-Object Optical and Near-infrared Spectrograph (MOONS)\cite{Taylor2018}, the Gemini Infrared Multi-Object Spectrograph (GIRMOS)\cite{Sivanandam2018}, the Extremely Large Telescope (ELT) High Angular Resolution Monolithic Optical and Near-infrared Integral field spectrograph (HARMONI)\cite{Thatte2010}, the Thirty Meter Telescope (TMT) InfraRed Imaging Spectrograph (IRIS)\cite{Larkin2010}, and the Maunakea Spectroscopic Explorer (MSE)\cite{McConnachie2016}. These instruments are designed for a wide variety of scientific use cases, including understanding the spectra of early galaxies, the kinematic and chemical structure of galaxies at cosmic noon, the chemistry of resolved stellar populations, and characterization of exoplanets.\cite{Barton2010,Thatte2010,McConnachie2016,Taylor2018,Boker2022}

\begin{figure}[ht]
    \centering
    \includegraphics[width=0.9\linewidth]{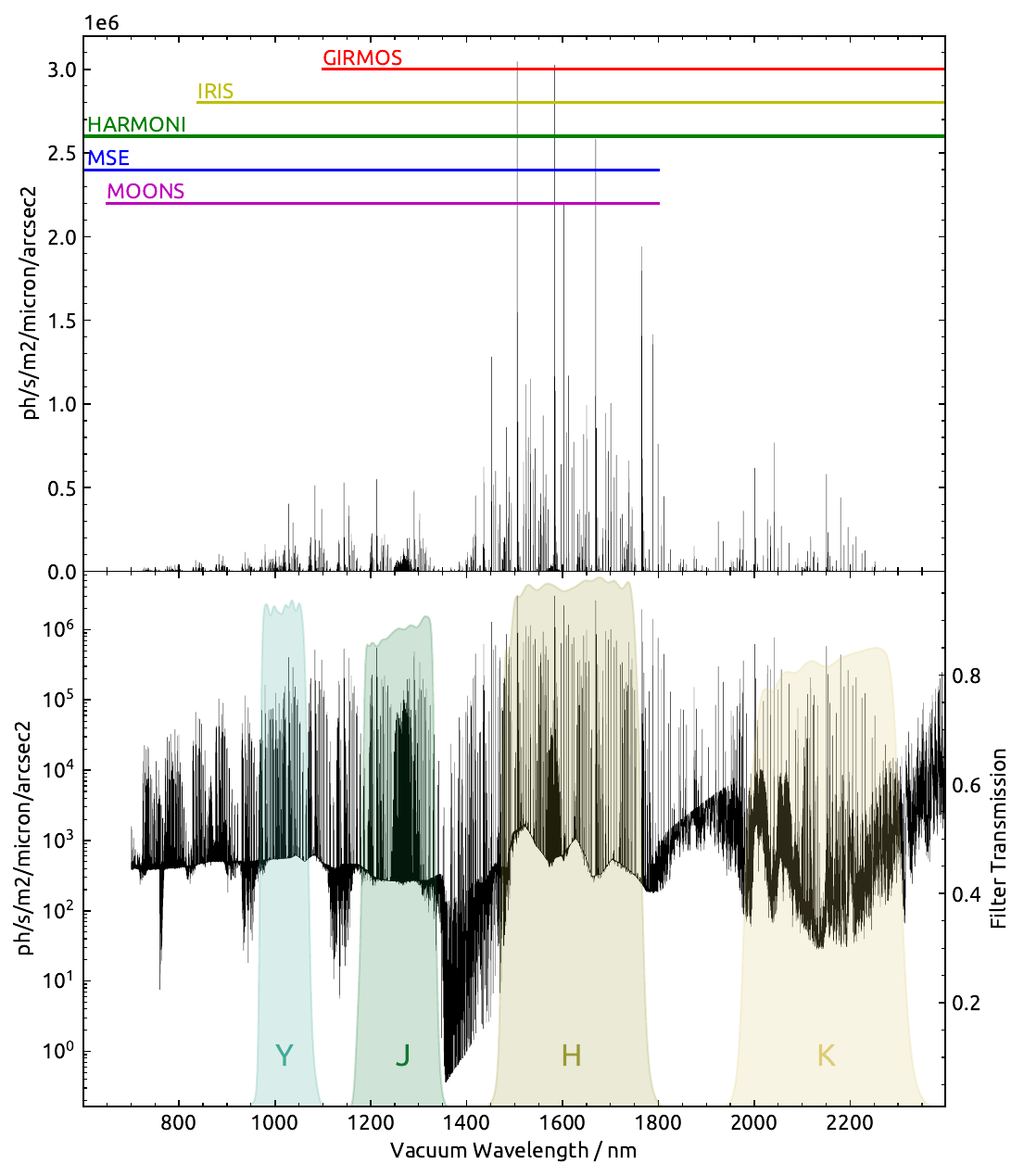}
    \caption{OH atmospheric emission lines, calculated with the ESO Skycalc model\cite{Noll2012,Jones2013}. The top panel shows the wavelength ranges of selected NIR spectrographs on top of the Meinel bands (linear scale). The bottom panel shows the total sky background (log scale) as seen at the VLT along with the transmission of selected VLT HAWK-I filters\cite{Rodrigo2012,Rodrigo2020} representative of typical NIR filters.}
    \label{fig:skycalc}
\end{figure}

Despite their large apertures, relative flexibility, and maintainability, ground-based telescopes have the disadvantage of looking through the Earth's atmosphere. One of the numerous challenges posed by the atmosphere is the OH airglow, which is composed of several rotation-vibration emission bands of the OH molecule\cite{Meinel1950}, shown in Figure \ref{fig:skycalc}. The OH bands span a large portion of the NIR, from 0.7--2.4~\um{}, with the brightest around 1.6~\um{} in the H-band, within the wavelength ranges of many of the upcoming NIR spectrographs. Much of the NIR spectroscopic instrumentation in development requires targeting sources fainter than $\rm{H}=25$~AB~mag, whereas the sky continuum between the OH lines is around 19--20~AB~mag~as$^{-2}$ in the J- and H-bands, and the lines themselves can be up to $10^3$ times brighter still.\cite{Sullivan2012,Oliva1992,Maihara1993} Left unchecked, saturated lines can lead to long-term persistence issues, causing difficulties associated with calibration and sensitivity\cite{Chapin2024}.

OH lines are also highly variable, with flux varying around 10\% on timescales of a few minutes, as well as variations up to 50\% over the course of a night.\cite{Ramsay1992} While the strengths of individual lines tend to be correlated within vibrational bands, the relative strengths of any two lines will not remain constant.\cite{Davies2007} As a result, it is necessary to take frequent sky exposures to be able to effectively subtract the sky spectrum from the target.

Even with frequent exposures, variability over the course of a single science exposure can lead to imperfect subtraction.\cite{Davies2007} For multi-object and integral field spectrographs, it is necessary to obtain well-measured sky spectra using either simultaneous sky observations with dedicated fibers or frequent offset exposures. To remove the sky spectrum from the target with minimal residuals, it is then necessary to have a good subtraction scheme, which will often use the full range of OH lines to inform the subtraction (e.g., Refs. \citenum{Wild2005,Davies2007,Soto2016}). As a result of the brightness and variability of the OH lines, the lines must be either measured to high precision for sky subtraction, or they must be removed completely to prevent them from impacting the target spectrum in the first place.\cite{Ellis2008}

Current strategies to remove OH lines from spectra tend to rely on customized optical components. For example, specially-selected narrowband filters (FWHM $<1$~nm) can probe the gaps between OH bands at the cost of removing most of the NIR spectrum, and are best suited for highly specific use cases such as the DAzLE instrument which is tuned for the detection of Lyman-$\alpha$ at $z>7$\cite{Horton2004}. High dispersion masking samples the continuum between individual lines by physically blocking the highly-dispersed light at predetermined positions, then recombining the light before sending it to the science instrument\cite{Maihara1993a}. Fiber Bragg gratings avoid the issue of grating scattering by using interference filters to remove specified wavelengths of light prior to dispersion \cite{Bland-Hawthorn2004}. While these methods are all successful at preventing the emission lines from overwhelming light from the target, they all require specialized optical components and have the limitation of removing all information at each wavelength that is blocked. By doing so, the masked OH lines cannot be used to inform subtraction schemes for the hundreds of fainter OH lines which cover the NIR.

When the OH lines are left in the spectrum, their brightness and variability necessitate the use of short exposure times to prevent their saturation and enable sky subtraction. Post-processing tools can recover useful information from saturated pixels,\cite{Finger2008} but this does not prevent the issues caused by saturation, so short exposures may still be preferred. In some cases, combining several short exposures has the disadvantage of increased detector read noise as compared to a single, longer exposure, increasing with the square root of the number of exposures. The total signal-to-noise ratio (SNR) for a background-subtracted pixel follows \begin{equation}\mathrm{SNR} = \frac{S_0nt}{\sqrt{S_0nt + 2(nt\Sigma S_N + nR^2)}},\label{eq:snr}\end{equation} where $S_0$ is the source flux, $n$ is the number of stacked exposures, $t$ is the exposure length, $S_N$ are the sources of background flux, and $R$ is the RMS read noise. For observations where read noise is a major contributor to the total noise, the number of stacked exposures increases noise faster than the length of the exposures. As a result, longer exposures are favoured for faint targets that may be accessible in spectral regions with fainter background, but this cannot be achieved without saturation or loss of information about the line variability. In practice, we find that modern spectrographs are rarely read-noise dominated, and so this is less of a concern for major telescopes.

\begin{figure}
    \centering
    \includegraphics[width=0.9\linewidth]{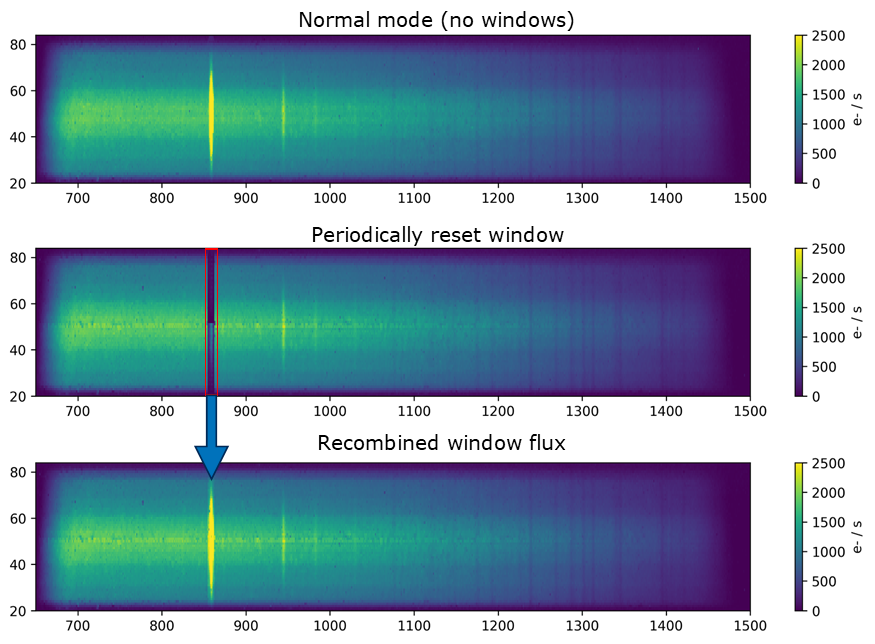}
    \caption{The top image is a spectrum with emission lines observed using a normal, no-window scheme. The middle image shows the placement of a window (outlined in red) on the same spectrum which is periodically reset; the data from this window are then recombined back into the image to create the bottom image. In a working demonstration of the concept, the top and bottom images should be close to identical.}
    \label{fig:ODESSA}
\end{figure}

In this paper, we present an on-detector method for preventing saturation of the OH lines during long exposures without removing them altogether, while simultaneously measuring their variability at a higher cadence, thus providing additional information for precise sky subtraction schemes. This method uses the ``guide window'' mode of a Teledyne HAWAII-2RG detector to selectively reset pixels before they reach saturation, while continuing integration on the rest of the detector. By carefully selecting the region of pixels which will be reset at a faster rate than the rest of the detector, which we refer to here as Bright Region Reset (BRR) mode, individual sky lines at any position on the detector can be suppressed. In addition, the charge in these window pixels is still measured during integration so that the full image can be reconstructed, including the flux within the windows. The high-frequency readout of these windows also results in higher signal-to-noise measurements of the lines themselves compared to a conventional integration. An illustration of this process is shown in Figure \ref{fig:ODESSA}.

While guide windows on HxRG detectors have been explored, they remain underutilized on science instruments in our opinion, and they have mostly been used in the context of fast measurements of guide stars\cite{Albert2005,Bezawada2006,Boss2009,Young2012,Smith2012}. A concept similar to BRR has been implemented on the Carnegie Astrometric Planet Search Cameras (CAPSCam), the Gemini South Adaptive-Optics Imager (GSAOI), and the ELT's Multi-AO Imaging Camera for Deep Observations (MICADO) by using resets of windows to prevent saturation of bright objects in the imaging field\cite{Boss2009,Young2012,Bezawada2024}. However, analysis of the performance is limited, and to our knowledge the applications have not been extended to spectra. A similar row-wise resetting scheme has been demonstrated to increase dynamic range on CMOS detectors, but has not been implemented in the NIR nor in spectra\cite{Wocial2022}. We note that the BRR method should be applicable to any detector system for which windowed readout/reset mode is possible.

In Section \ref{sec:setup}, we describe the experimental design and setup necessary for our on-telescope demonstration of this process for suppressing bright emission lines. Section \ref{sec:data} describes the data and reduction used in the analysis. Section \ref{sec:results} presents the results of the analysis as well as some limitations imposed by our setup. In Section \ref{sec:future}, we discuss how BRR might be implemented on future spectrographs and what results can be expected. Section \ref{sec:conclusions} summarizes our work.

\section{Experimental Setup}
\label{sec:setup}

Herzberg Astronomy and Astrophysics (HAA) possesses a 5~\um{}--cutoff Teledyne HAWAII-2RG which was a flight spare for the JWST Fine Guidance Sensor. This detector is not ideal for our application because of its extended long-wavelength sensitivity, making it highly susceptible to background thermal radiation. However, we used it due to lack of availability of a suitable shorter-wavelength (e.g., 2.5~\um{}--cutoff) detector. We pair this detector with the ARC Gen-4 controller used by Chapin et al.\cite{Chapin2022} and the McKellar Spectrograph at the 1.2~m telescope of the Dominion Astrophysical Observatory (DAO) to demonstrate the BRR concept.

\subsection{Spectrograph Setup}

The McKellar Spectrograph is situated at the Coud\'e focus of the 1.2-m telescope at the DAO. Its selection of grating configurations has produced a large number of high-resolution stellar spectra for spectral classification and radial velocity measurements\cite{Richardson1968}. Though the gratings are mostly optimized for visible wavelengths, the grating we selected (600~lines~mm$^{-1}$, blaze 700~nm, R${\sim}20,000$) has sufficient throughput at 1~\um{} to use with our detector (${\sim}60\%$ relative to 700~nm). This was determined using a SITe-4 CCD mounted to the spectrograph: a broadband light source was imaged through narrowband filters centred at 700, 850, and 1000 nm for several gratings, and source intensity at the entrance slit was measured with a calibrated photodiode.

\begin{figure}
    \centering
    \includegraphics[width=0.9\linewidth]{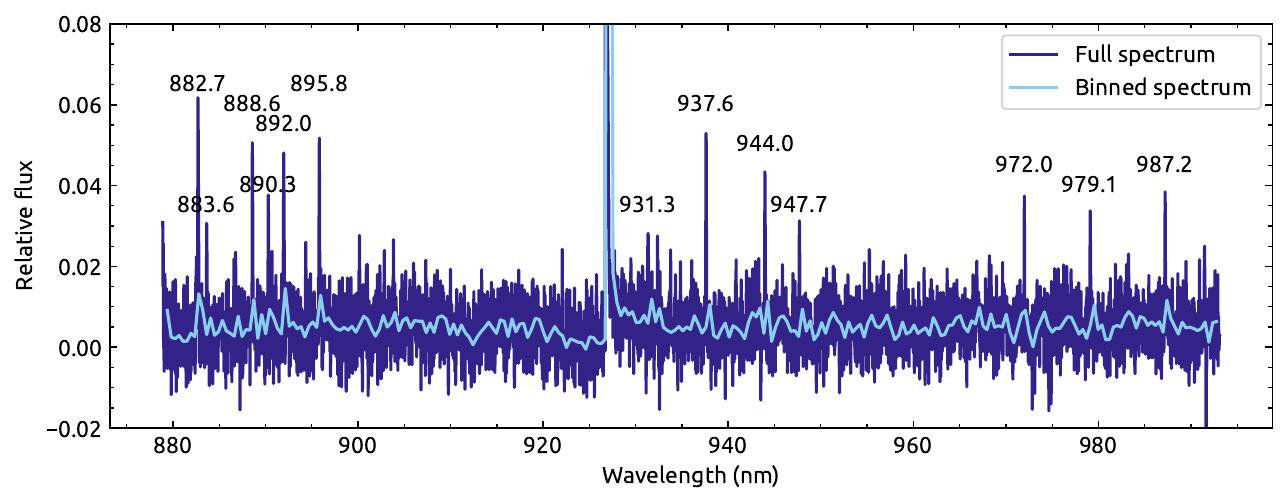}
    \caption{Observed night sky spectrum at the McKellar Spectrograph, with known sky lines labelled with their wavelengths. The large spike around 927~nm is a bad row on the CCD. The 15-px binned spectrum is also overlaid. Low throughput of our system led us to use arc lamps instead of sky lines for our demonstration.}
    \label{fig:mckellar-sky}
\end{figure}

To determine which OH lines visible to the McKellar would be suitable for a demostration of BRR, we carried out a series of observations of the sky using this grating and the CCD that is typically used to observe with this spectrograph. We found that a combined 12 $\times$ 900-second exposures, shown in Figure \ref{fig:mckellar-sky}, could barely distinguish OH lines from the background noise between 880 and 980~nm, which includes the (7, 3) and (8, 4) OH vibrational bands. The relatively small aperture of the telescope and a deteriorated primary mirror coating at the time of the experiment limited the amount of light from the sky which reached the detector (the mirror has since been recoated). Optimizing the configuration of the spectrograph was therefore not considered likely to result in significantly brighter lines.

\begin{figure}
    \centering
    \includegraphics[width=0.8\linewidth]{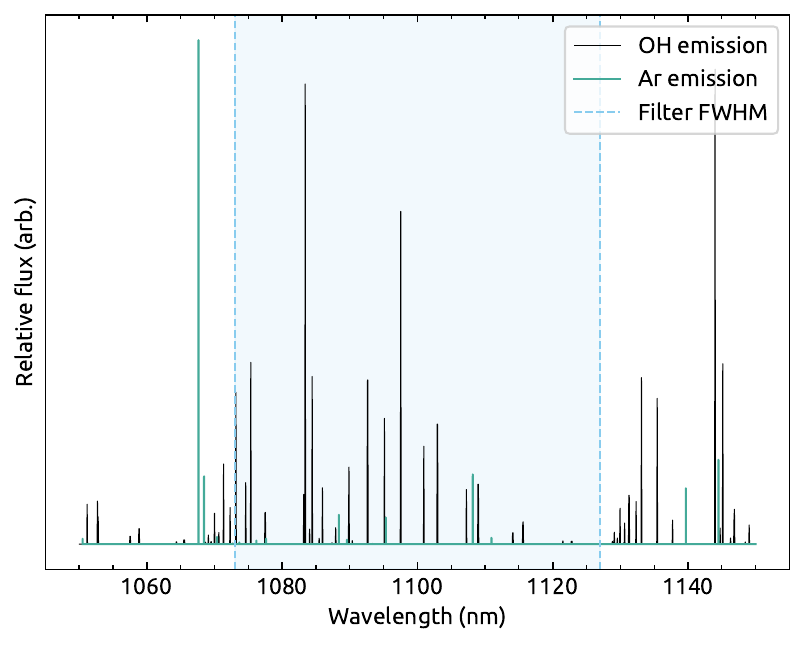}
    \caption{Locations and relative strengths of OH and Ar emission lines, along with the FWHM range of our chosen narrowband filter. We use the 1108 and 1111~nm Ar lines which lie within the (5, 2) OH vibrational band.}
    \label{fig:lines-filter}
\end{figure}

Given the absence of bright sky lines that could be observed with the McKellar, we instead decided to use a different method to demonstrate this technique, using arc lamps to create artificial sky lines. Because BRR is useful for resetting any emission line which may saturate the detector quickly, a demonstration with an artificial sky spectrum is just as useful as a demonstration with real sky spectra. The 1.2-m dome has an FeAr calibration lamp with strong Ar emission lines in the vicinity of 1~\um{}. Our selection of filters (see Section \ref{sec:cryostat}) ultimately resulted in our selection of the 1108~nm and 1111~nm Ar lines to use as mock-OH lines. These lines are a good proxy as their wavelengths lie within the (5, 2) vibrational band of OH, as shown in Figure \ref{fig:lines-filter}.

\begin{figure}
    \centering
    \includegraphics[width=0.9\linewidth]{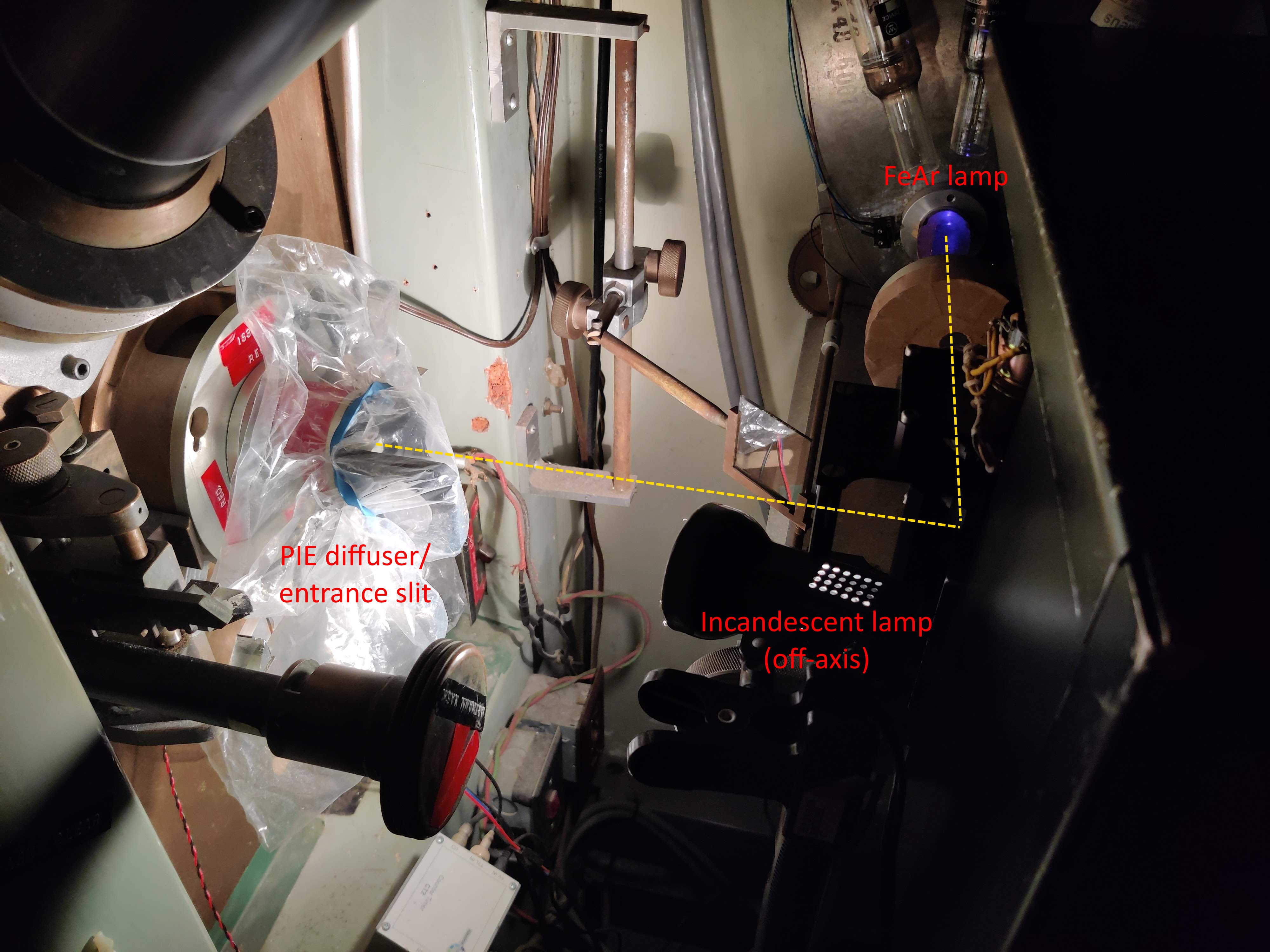}
    \caption{The entrance slit setup used for the light source. The PIE diffuser is on the left, covering the entrance slit. The clip-on incandescent lamp is in the center, and the arc lamp is in the top right. The arc light is redirected into the beam by a flat mirror. The optical path is indicated by the yellow dashed line.}
    \label{fig:PIE-diffuser}
\end{figure}

To create a light source which mimics some astronomical spectrum contaminated by sky lines, we pair the arc lamp with an incandescent lamp, representing some arbitrary target continuum spectrum. While the McKellar Spectrograph is equipped with both lamps, it does not have a way to place both sources in the optical path at the same time. Instead of using the McKellar incandescent lamp, we attached a clip-on lamp to a tripod standing just outside the entrance slit, aiming it directly at the slit while the arc lamp was activated. To diffuse the incandescent light into the beam, we used a double-layered PIE (Plastic-bag Improvised Elastic-secured) diffuser---this setup is shown in Figure \ref{fig:PIE-diffuser}. The PIE diffuser scatters enough light to redirect the incandescent light into the beam, but does not appreciably attenuate the light from the arc lamp. The resulting spectrum has a broad continuum truncated by the edges of the beam, with the two strong Ar emission lines clearly visible above the continuum, as was seen in Figure \ref{fig:ODESSA}, for example.

\subsection{Cryostat Setup}
\label{sec:cryostat}

The experiments of Chapin et al.\cite{Chapin2022} mounted an H2RG in a $^4$He closed-cycle cryostat. However, this cryostat does not fit on the mounts at the McKellar Spectrograph, so we transferred this detector to a smaller, liquid nitrogen (LN2)--cooled cryostat. The detector is mounted in its JWST flight package, which is attached directly to the cryostat cold plate.

The LN2 cannot passively cool the detector below 77K, so to reduce dark current and thermal background in the 5~\um{}--cutoff detector we attached a vacuum pump to the LN2 tank that abuts the cold plate. A pressure of a few Torr was able to lower the cold plate temperature to around 55K. This resulted in a detector temperature around 65K during operation.

\begin{figure}
    \centering
    \includegraphics[width=0.9\linewidth]{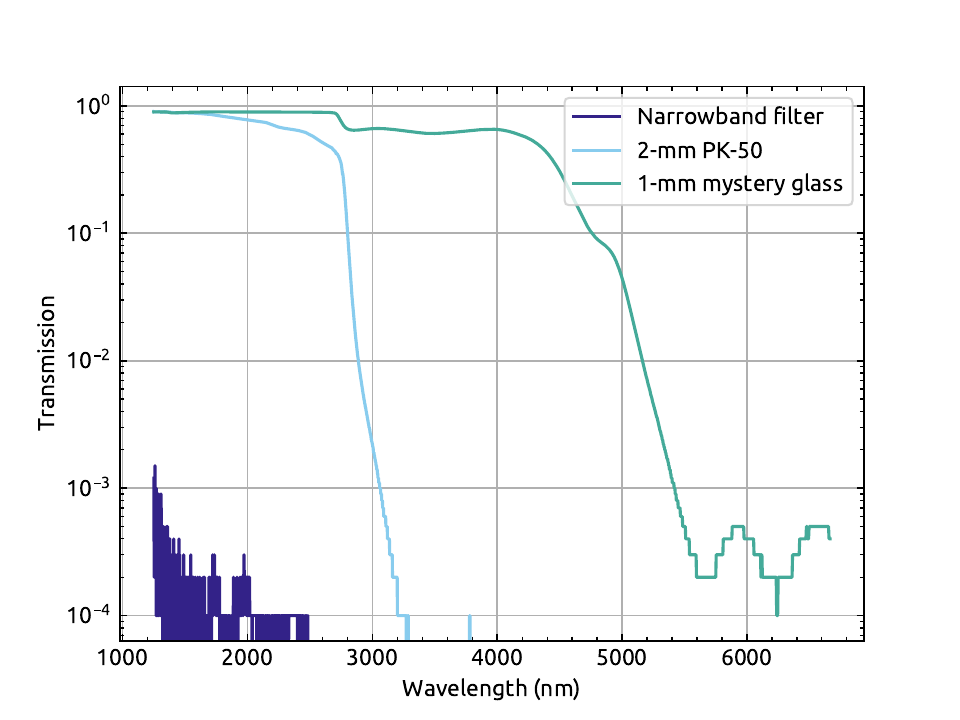}
    \caption{Measured filter transmission spectra, to a precision of $10^{-4}$. The narrowband filter is the 1100~nm filter from Andover, and ``PK-50'' and ``mystery glass'' are IR blocking filters from HAA.}
    \label{fig:transmission}
\end{figure}

This detector is sensitive to 5~\um{} radiation from the 300K Coud\'e room, so we include several filters to prevent immediate saturation from infrared (IR) background photons. To allow us to image arc lines, we searched for a narrowband filter which spanned a few available Ar lines. The one we selected is a 1100~nm centre wavelength, 54~nm FWHM surplus filter from Andover Corporation, which encompasses the 1108~nm and 1111~nm lines mentioned above. To further reduce the thermal background, we also included a 2~mm thick piece of IR-blocking PK-50 glass, as well as a 1~mm thick unknown glass which was found to reduce the total background by about half. The transmission of each of these filters from 1.25--6.67~\um{} was measured using a Spectrum Two\texttrademark{} IR spectrometer at the University of Victoria. The resulting transmission spectrum of each filter to a precision of $10^{-4}$ is shown in Figure \ref{fig:transmission}. We estimate the total transmission of the combined filter stack is less than $10^{-9}$ at 5~\um{}.

\begin{figure}
    \centering
    \includegraphics[width=0.9\linewidth]{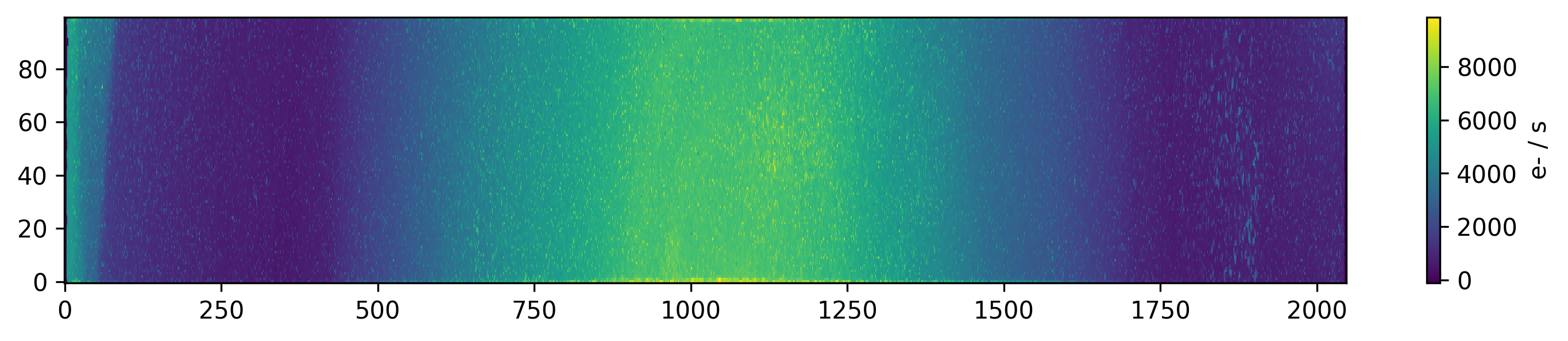}
    \caption{Typical background seen in dark images using our setup.}
    \label{fig:bkgnd}
\end{figure}

Due to the small available volume within the cryostat, the filter holder is mounted to the radiation shield instead of the cold plate, which only reached temperatures around 85K during operation. This meant that, despite the efforts to reduce the background as much as possible, there was still a significant glow coming from the filters themselves in addition to any remaining external radiation. The background, shown in Figure \ref{fig:bkgnd}, posed significant challenges to our experiment, as it would saturate the centre region of the detector in around 10~seconds. The tests we carried out thus had to be designed around this restriction.

\subsection{Detector Readout}

\begin{figure}
    \centering
    \includegraphics[width=0.9\linewidth]{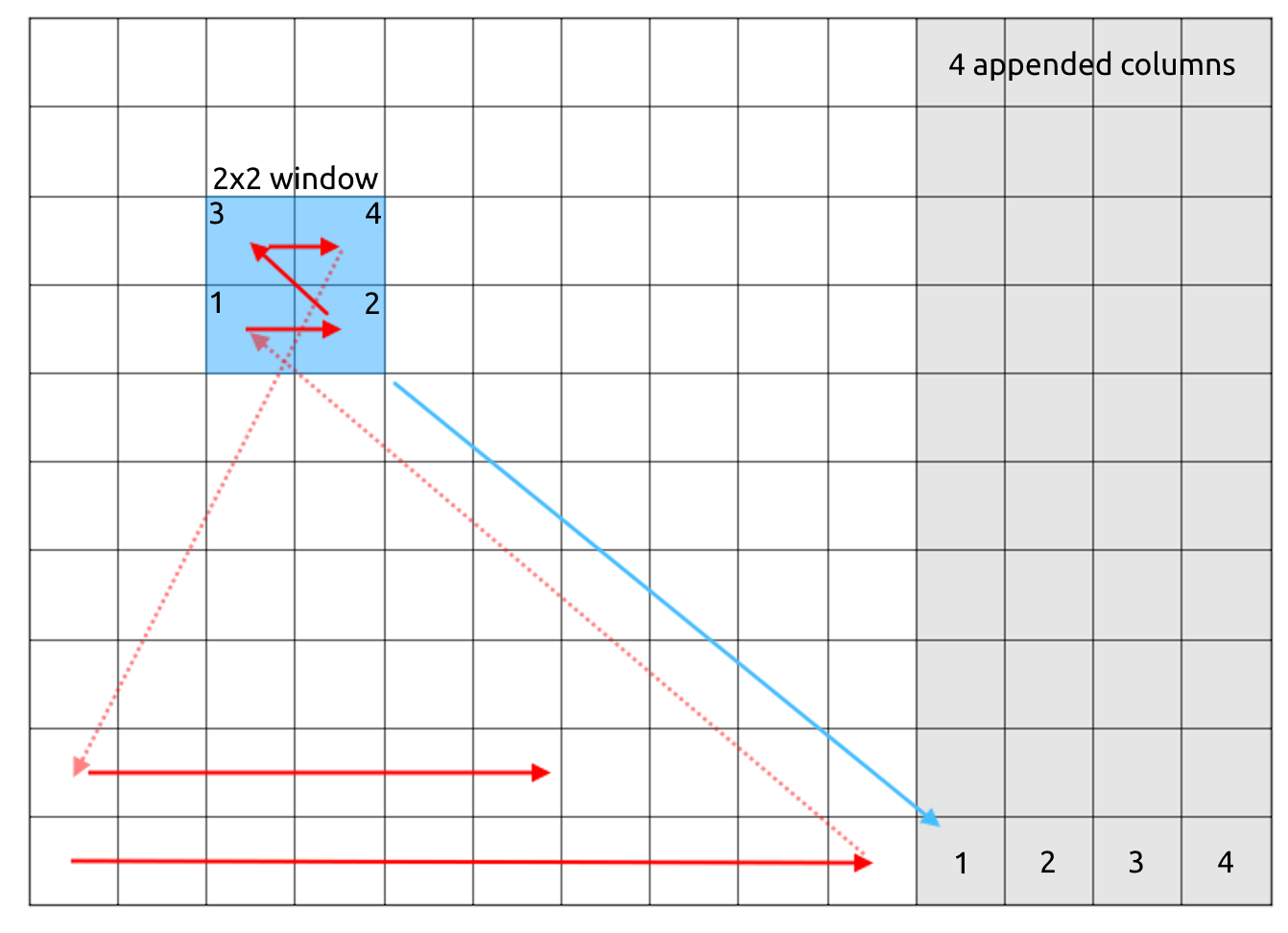}
    \caption{Illustration of how windows are read out with a small $2\times2$ window within the larger frame. Red arrows represent the order in which pixels are visited. Window pixels are appended to the end of a row according to their visit order, indicated by the numbers.}
    \label{fig:guide-window}
\end{figure}

The BRR readout of the detector is adapted from the On-Detector Guide Window readout mode of Chapin et al.\cite{Chapin2022} using an ARC Gen-4 controller. BRR implements up-the-ramp (UTR) sampling for the full frame, while additionally sampling an entire guide window at the end of each row. It then appends the values of the entire window to the end of the row. Each window can be optionally reset when it is visited as specified by a reset period for each window. If $n$ windows have been defined in the exposure, the windows are visited in succession so that each window is read once every $n$ full-frame rows. An illustration is shown in Figure \ref{fig:guide-window}, where the grid represents an arbitrary region of the detector being read out. The clocking scheme is mostly the same as in Chapin et al.\cite{Chapin2022}, with the addition of timing delays during window reads to ensure the row period remains the same regardless of window size. These delays set the maximum window size to 32~rows $\times$ 16~columns for our software. Larger windows are achievable, but would require longer delays.

Configurable parameters include the number of windows, the number of UTR reads to include after an initial reset, how often to reset the full frame, how often to reset the window(s), the  size of the window(s), and the location(s) of the window(s). In the case of multiple windows, each window can have an independent reset rate and location, but they must all be the same dimensions.

The existing cryostat wiring restricted us to transferring data through a single video channel (compared to the maximum 32-channel output), limiting the readout rate to around 40~seconds for a single $2048\times2048$ pixel frame---a factor of a few times longer than it takes to reach saturation from the thermal background. In our tests, we instead choose to skip over the first $\sim$1000 rows and read only the central 100 rows which contain the spectral information. This results in a frame period of 2.7~s, which was generally long enough to obtain 4 UTR samples per pixel before saturation.

\subsection{Gain and Read Noise Characterization}
\label{sec:characterization}

In order to understand the characteristics of our detector and provide uncertainty estimates, we measure the electron conversion gain and read noise. The typical method for measuring gain is by taking the slope of a mean--variance curve (commonly called a photon transfer curve) created with flat field images. The thermal background in our images makes uniform flat field images across a range of flux levels not feasible to obtain, but it allows individual images to be split into $4\times4$ regions of pixels with uniform mean fluxes. We take fifty 4-sample UTR ``dark'' images and take samples from the area of the detector receiving the majority of the background flux. Within each $4\times4$ region, we select either the first, second, or third read from each of the 50 ramps and use this sample to create a mean--variance curve. The slope of this curve is measured using ordinary least squares. Assuming each pixel across the detector has approximately the same gain, the average gain of the detector in e$^-$~ADU$^{-1}$ is $1/{\rm{median}}([a_1,a_2,...])$, where $a_i$ is the measured slope of one $4\times4$ region. We measure an average gain of $3.0\pm0.6~\rm{e}^-~\rm{ADU}^{-1}$, which is consistent with the expected gain for this detector.\cite{Chapin2022}

The mean--variance method can also be used to measure read noise, but it requires finely-sampled data in the low-flux regime.
We therefore estimate the read noise by finding the variance of correlated double samples (CDS) of the reference pixels across a sequence of exposures. Light-sensitive pixels are excluded due to the shot noise of the background even with the shortest possible exposure time. We find the median CDS read noise to be $34\pm7~\rm{e}^-$, corresponding to a single-sample read noise of $24\pm5~\rm{e}^-$. This measurement is somewhat high relative to other HxRG detectors\cite{Rauscher2007,Beletic2008,Artigau2018}, and may be a result of our choice of controller. We emphasize, however, that this discrepancy does not affect our main conclusions regarding the performance of the BRR scheme.

\section{Data Collection and Reduction}
\label{sec:data}

A successful demonstration of on-detector sky line suppression will show whether or not periodically resetting bright emission lines significantly changes an observed spectrum compared to a more standard readout mode. In particular, we want to show whether systematic effects caused by windows significantly affect the features of the final 1-D coadded spectra. If they do not, then it may be useful to use this method in actual observations that are dominated by sky emission lines.

We choose to compare the BRR method to the conventional readout method in two ways. First, we periodically reset an entire spectral line and compare the resulting image to one with no windows; and second, we reset only the lower half of the spectral line, and separately analyze and compare the upper and lower halves. The latter method has the advantage that we observe the BRR and conventional methods at exactly the same time, thus removing some systematic uncertainties such as varying thermal background between exposures. An undiagnosed bug required our windows to be shorter than the full height of the spectrum, so in order to reset an entire emission line, we placed two 32-row $\times$ 8-column windows adjacent to each other to cover the whole line.

In order to build up a large number of observations for statistical analysis, we took several hours of images with different window configurations. In addition to the no-window, half window, and full window configurations, we included equivalent images in which the window was being read only (and not reset), so that we might be able to distinguish between any features in the images caused by only reading a window versus reading and resetting it. For each of these window configurations, we took dark and arc-plus-incandescent exposures as well as arc-only and incandescent-only exposures.

To prevent the slowly-changing background flux from creating a time-dependent signal during dark subtraction, we interleaved dark images throughout the imaging sequences. Images were divided into sequences of 20 like images in a row, taking about 5 minutes per sequence. Each group of sequences with the same window configuration is preceded by a dark sequence, and the groups of sequences are repeated several times. A summary of the data collected is given in Table \ref{tab:data}. Due to an undiagnosed instability in the analog-to-digital converter (ADC) offset, one of the groups of no-window sequences was discarded. During the data collection, the detector temperature remained around 63K while the radiation shield was around 86K.

\begin{table}
    \centering
    \caption{Summary of observations. Sequence type ``spectrum'' refers to an image containing both the arc and incandescent light; ``arc'' is arc only, and ``flat'' is incandescent only. Eight corrupted images were discarded, as well as one repetition of the ``No window'' sequences. Each exposure is approximately 8~seconds.}
    \label{tab:data}
    \begin{tikzpicture}[overlay, remember picture]
        \draw[decorate,decoration={brace,amplitude=10pt}] (-0.2,-3.4) -- (-0.2,4.7) node[midway,xshift=-20pt]{\rotatebox{90}{\centering Repeated 4 times}};
    \end{tikzpicture}
    \begin{tabular}{|c|c|} \hline
        Window configuration & Sequence \\ \hline
        \multirow{4}{*}{No window} & 20 dark \\ \cline{2-2}
        & 20 flat \\ \cline{2-2}
        & 20 spectrum \\ \cline{2-2}
        & 20 arc \\ \hline
        \multirow{3}{*}{Half 1108~nm line, read only} & 20 dark \\ \cline{2-2}
        & 20 spectrum \\ \cline{2-2}
        & 20 arc \\ \hline
        \multirow{3}{*}{Half 1108~nm line, reset every 50 rows} & 20 dark \\ \cline{2-2}
        & 20 spectrum \\ \cline{2-2}
        & 20 arc \\ \hline
        \multirow{3}{*}{Full 1108~nm line, read only} & 20 dark \\ \cline{2-2}
        & 20 spectrum \\ \cline{2-2}
        & 20 arc \\ \hline
        \multirow{3}{*}{Full 1108~nm line, reset every 50 rows} & 20 dark \\ \cline{2-2}
        & 20 spectrum \\ \cline{2-2}
        & 20 arc \\ \hline
        \addlinespace[2mm] \hline
        \multirow{3}{*}{\makecell{1108~nm line, reset every 50 rows; \\ 1111~nm line, reset every 100 rows}} & 20 dark \\ \cline{2-2}
        & 20 spectrum \\ \cline{2-2}
        & 20 arc \\ \hline
    \end{tabular}
\end{table}

To create 1D spectra from the 3D UTR FITS files in which each image is stored, we again adapt the reduction steps from Chapin et al.\cite{Chapin2022} We start by performing a reference pixel subtraction. For each reference pixel, we fit a line to each pixel as a function of read number and calculate the residuals. For each row in a frame, we then subtract the median residual of its eight reference pixels.

We create a master dark image for each of the consecutive image sequences with the same window configuration. This is done by taking the median of a series of reference-subtracted dark data cubes with window reads still appended to each row. The resulting master dark is subtracted directly from a reference-subtracted data cube. Bad pixels are flagged using an iterative sigma-clipping process on both the flux and reset values of the individual dark images. These flagged pixels are excluded from the remainder of the analysis. Around 27\% of pixels were masked in the final data runs.

After dark subtraction, we separate the window reads from the full frame reads and create 2D flux images for each by fitting the slope value of each pixel using ordinary least squares. The variance of the slope is calculated using the method of Refs. \citenum{Rauscher2007} and \citenum{Robberto2009}, which accounts for the correlation between UTR samples in each pixel:
\begin{equation}
    \sigma_f^2 = \frac{6}{5} \frac{n^2+1}{n(n^2-1)} \frac{f}{t} + 12 \frac{1}{n(n^2-1)} \frac{\sigma_{RN}^2}{t^2},\label{eq:var}
\end{equation}
where $f$ is the measured slope, $n$ is the number of samples in the ramp, $t$ is the integration time, and $\sigma_{RN}$ is the single-sample read noise. To report the flux and uncertainty in e$^-$~s$^{-1}$, we use the average gain of 3.0~e$^-$~ADU$^{-1}$ and read noise of 24~e$^-$ measured in Section~\ref{sec:characterization}.

In the case of a window which has been reset during the exposure, a new ramp is fit after every reset. A total flux and uncertainty for the pixel for one exposure is calculated by finding the weighted mean of each individual fitted flux,
\begin{equation}
    f_{exp} = \frac{\Sigma f_i~\sigma^{-2}_{f,i}}{\Sigma \sigma^{-2}_{f,i}},
\end{equation}
where $f_i$ is the flux of one ramp between resets, and $\sigma_{f,i}$ is the uncertainty in the measured flux of that ramp. Since the flux of the full frame and window(s) of an exposure have been calculated separately, we then recombine the window flux into its appropriate location in the full frame, creating a single flux image using all the data collected.

To convert 2D flux images into 1D spectra, we use a weighted mean extraction with the row weights based on the incandescent-only images. Uncertainty spectra are extracted from a variance image using the same weights. Wavelength solutions are found from the extracted arc spectra by fitting a Gaussian to each of the 1108.19~nm and 1110.95~nm Ar emission lines, and fitting a linear relationship between wavelength and pixel.

After a group of like images has been extracted into 1D spectra, they are aligned onto a common wavelength grid corresponding to the range and twice the pixel resolution of the detector: 1101--1128~nm in 0.016~nm increments. This is done using the \texttt{specutils} Python package's \texttt{LinearInterpolatedResampler}, which resamples both the flux and uncertainty of a given spectrum onto a new grid. Finally, spectra are coadded using an inverse variance weighted mean and normalized by their mean. The final coadded error spectra represent the total RMS noise per wavelength resolution element.

\section{Results}
\label{sec:results}

\begin{figure}
    \centering
    \includegraphics[width=0.9\linewidth]{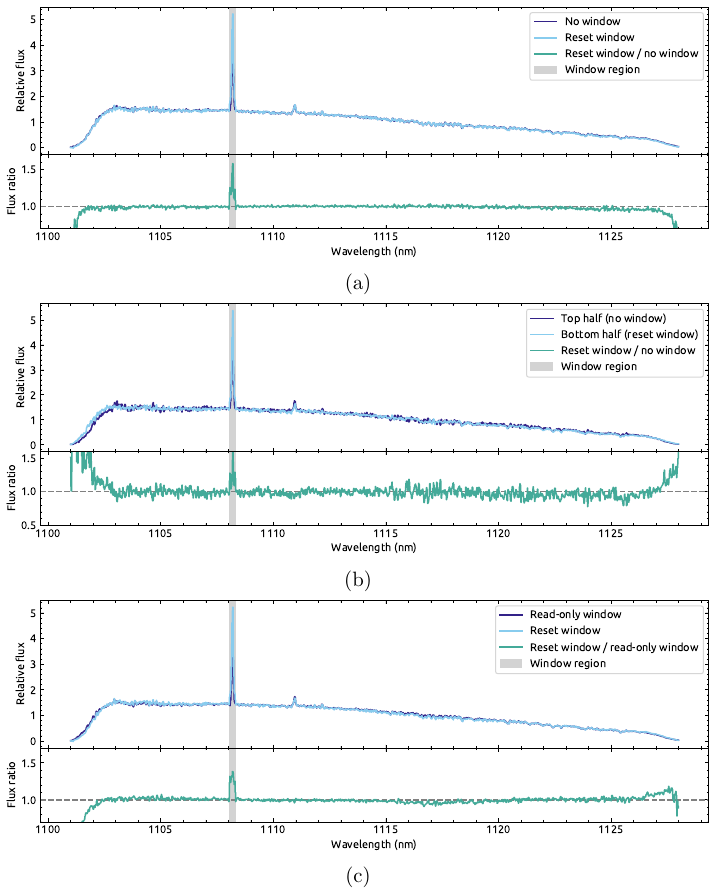}
    \caption{(a) Comparison of spectra using no-window mode and BRR mode. The bottom panel shows the ratio of the two (BRR / no-window mode). (b) Same as (a), but instead comparing the two portions of the half-window-reset mode. (c) Same as (a), but instead comparing a read-only window to a reset window. The overall shapes of the spectra are unaffected by BRR, while flux is increased in regions containing a window which has been periodically reset. The increased flux within the windows is explained in Section \ref{sec:limitations}.}
    \label{fig:flux-comparison}
\end{figure}

The final spectra produced by the typical no-window exposures and the BRR mode exposures are shown in Figure~\ref{fig:flux-comparison}(a). Immediately we can see that the overall shape of the spectrum is unaffected by the use of the BRR mode. To aid in uncovering any differences, we also show the ratio of the two modes (BRR / no-window). The ratio reveals that the BRR mode results in a slightly brighter emission line, for reasons discussed in Section \ref{sec:limitations}. On either end of the spectrum, deviations of the ratio from unity are caused by slight shifts in the illuminated region of the detector across the two days of observations. These shifts are physical shifts of the detector in the optical path, so they can be safely ignored for our purposes.

Similarly, we can compare the two portions of the half-window-reset configuration described above to directly test the two modes on the same spectrum, without having to consider variations between exposures. The spectra are shown in Figure \ref{fig:flux-comparison}(b). As before, the only significant difference between the two spectra is the increased flux in the window being reset. A comparison between BRR mode and a window which is read but not reset yields the same results, seen in Figure \ref{fig:flux-comparison}(c), as does a comparison against the exposures that use four windows to reset both emission lines. The consistency between the spectra in each exposure mode (barring the increased flux in reset windows) shows that BRR mode does not introduce systematic errors to scientific observations.

\begin{figure}
    \centering
    \includegraphics[width=0.8\linewidth]{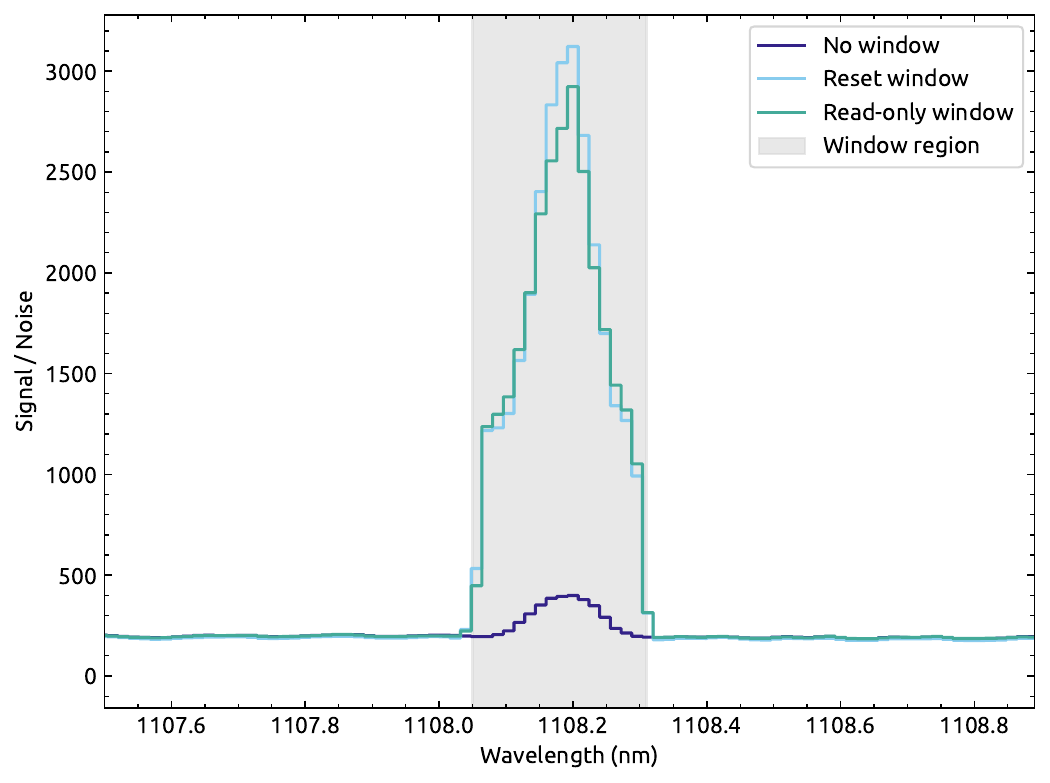}
    \caption{Comparison of SNR for the 1008~nm emission line for different observing modes. Operating a window significantly increases SNR due to the greater number of samples within the window.}
    \label{fig:SNR-window}
\end{figure}

The operation of windows during an exposure significantly increases the SNR within the window, as seen in Figure \ref{fig:SNR-window}. This change is due to the number of UTR samples to which a flux is fit.\cite{Artigau2018} Without a window, the variance of the slope is large due to a limited number of samples, while reading a window dramatically increases the number of samples (by a factor of 100 in this case). Periodically resetting the window increases the uncertainty somewhat, due to fitting slopes to several shorter ramps instead of one long ramp. In our experiment, this is offset by the increased measured signal within reset windows, though we expect this to be unique to our experiment (see Section \ref{sec:limitations}). Even with the increase in uncertainty, the BRR mode has a SNR significantly higher than the no-window mode, demonstrating the potential for the BRR mode to improve monitoring of quickly-varying sky lines. This, combined with the consistency between observing modes, indicates that the BRR readout scheme is both possible to implement and deserving of further study.

\subsection{Limitations}
\label{sec:limitations}

Our experimental setup led to some limitations in the nature of our tests, although they ultimately do not affect our conclusions. Figure \ref{fig:flux-comparison} showed that the measured flux within a window which is being reset is higher than the measured flux without any resets. The most likely cause for this effect is pixel nonlinearity. As charge accumulates within a pixel during an exposure, that charge decreases the effective bias voltage, resulting in a lower measured charge within the pixel\cite{Plazas2017}. Nonlinearity therefore becomes more significant at higher sections of pixel ramps.

Periodically resetting pixels in BRR mode means that those pixels never reach the nonlinear regime. As a result, the combined flux measurement for a BRR pixel is already quite close to the ideal, linear flux. Non-BRR pixels, on the other hand, behave nonlinearly as expected, and have a measured flux lower than a BRR pixel receiving the same amount of light.

We attempted a nonlinearity correction by fitting a second-order polynomial to each pixel's ADU ramp in a median dark image. The quadratic term of this polynomial is then added back into reference-subtracted ramps, so that the data follow the linear-only terms. Across the window, the median nonlinearity-corrected flux is $(42\pm11)$\% higher than the uncorrected flux; this is consistent with the increased flux seen in the window regions in Figure \ref{fig:flux-comparison} of $(27\pm15)$\% relative to exposures with no window. We conclude that the difference between spectra is due to the lack of a pixel nonlinearity correction.

Unfortunately, we cannot uniformly apply this nonlinearity correction across the entire detector. Variable nonlinearity across pixels requires each pixel to have its own correction, but the high background flux means that most pixels in our data saturate after four full-frame UTR reads for each image. This becomes a problem due to a feature dubbed the ``reset anomaly''\cite{Bezawada2006,Rauscher2007,Langevin2022,Chapin2022}. The anomaly is characterized by an abrupt increase in a pixel's apparent flux directly after being reset, before flattening to a more linear rate. The 2nd-order polynomial fit to the four samples in pixels featuring the reset anomaly overestimates the nonlinearity correction, producing wildly inaccurate ``corrected'' ADU values.

The anomaly is well-fit by a four-parameter model including a linear and an exponential term\cite{Rauscher2007}. Our restriction to only four samples per ramp makes this model of little use to us, though, and the model would need an additional parameter to account for nonlinearity at high fluence. An experiment that fits more samples up the ramp before saturation would be able to correct for both these features, so we expect that the systematic effect seen in our results is unique to our specific experiment. As a result, the nonlinearity effect discussed here does not impact our main conclusions regarding the utility of the BRR mode.

In addition to preventing a nonlinearity correction, the bright thermal background in our images limits the experiments that we can conduct. BRR is intended to allow exposures free from saturation of bright emission lines; however, our entire detector saturates soon after the emission lines saturate. As a result, we cannot directly characterize the behaviour of BRR in the context of long exposures on faint targets. To do so, it will be necessary to implement this procedure on a more controlled facility which has greater suppression of background light or a detector with a shorter wavelength cutoff.

\section{Applications for Premier Facilities}
\label{sec:future}

\begin{figure}
    \centering
    \includegraphics[width=0.9\linewidth]{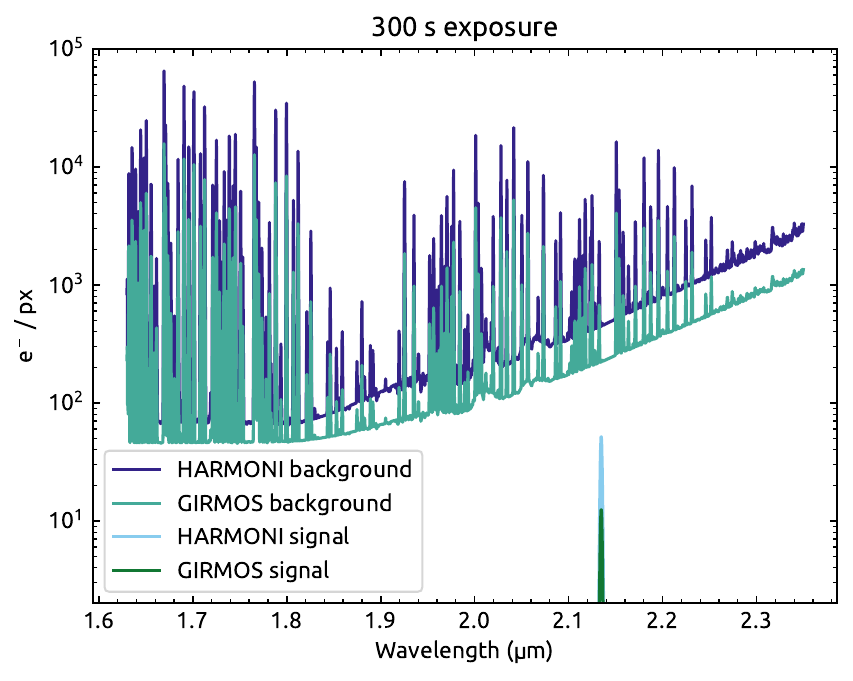}
    \caption{Simulated background and target spectrum as measured in one spaxel of GIRMOS and HARMONI in a 300~s exposure, mapped onto the pixel scale of GIRMOS (3.0~px/nm). Target signal is H$\alpha$ emission from a 40~M$_\odot$/yr, $z=2.3$ galaxy. HARMONI reaches saturation of the OH lines much more quickly than GIRMOS.}
    \label{fig:GIRMOS-spec}
\end{figure}

As an example of what advantages BRR could have on a major telescope, we use a preliminary version of the GIRMOS exposure time calculator to estimate the observed spectrum of a science target contaminated by the sky and instrument background\cite{Lamb2022}. GIRMOS is a new four-object integral field spectrograph under development for Gemini Observatory. In addition to its first-of-its-kind multi-object AO--assisted, multiplexed integral field capabilities, GIRMOS will be a pathfinder for multi-object spectroscopy on ELT-class instruments\cite{Sivanandam2018}. Figure \ref{fig:GIRMOS-spec} shows the simulated H$\alpha$ emission of a star-forming galaxy at $z=2.3$ with a star formation rate of 40~M$_\odot$/yr as seen by GIRMOS, along with the sky and thermal background for the instrument. The OH lines are relatively weak for GIRMOS due to its small 100~mas spaxel size, taking 30~minutes for the brightest OH lines to reach the 100~ke$^-$ full well of the H4RG detectors. However, assuming similar noise properties, HARMONI (the first-generation NIR integral field spectrograph for ELT, with a 39.3~m aperture and up to $30\times 60$~mas spaxel) reaches saturation in only 7~minutes. Seeing-limited spectrographs with larger fibers, such as MSE\cite{Hill2018}, may reach saturation in less than 3 minutes for the brightest lines.

\begin{figure}[ht]
    \centering
    \includegraphics[width=0.9\linewidth]{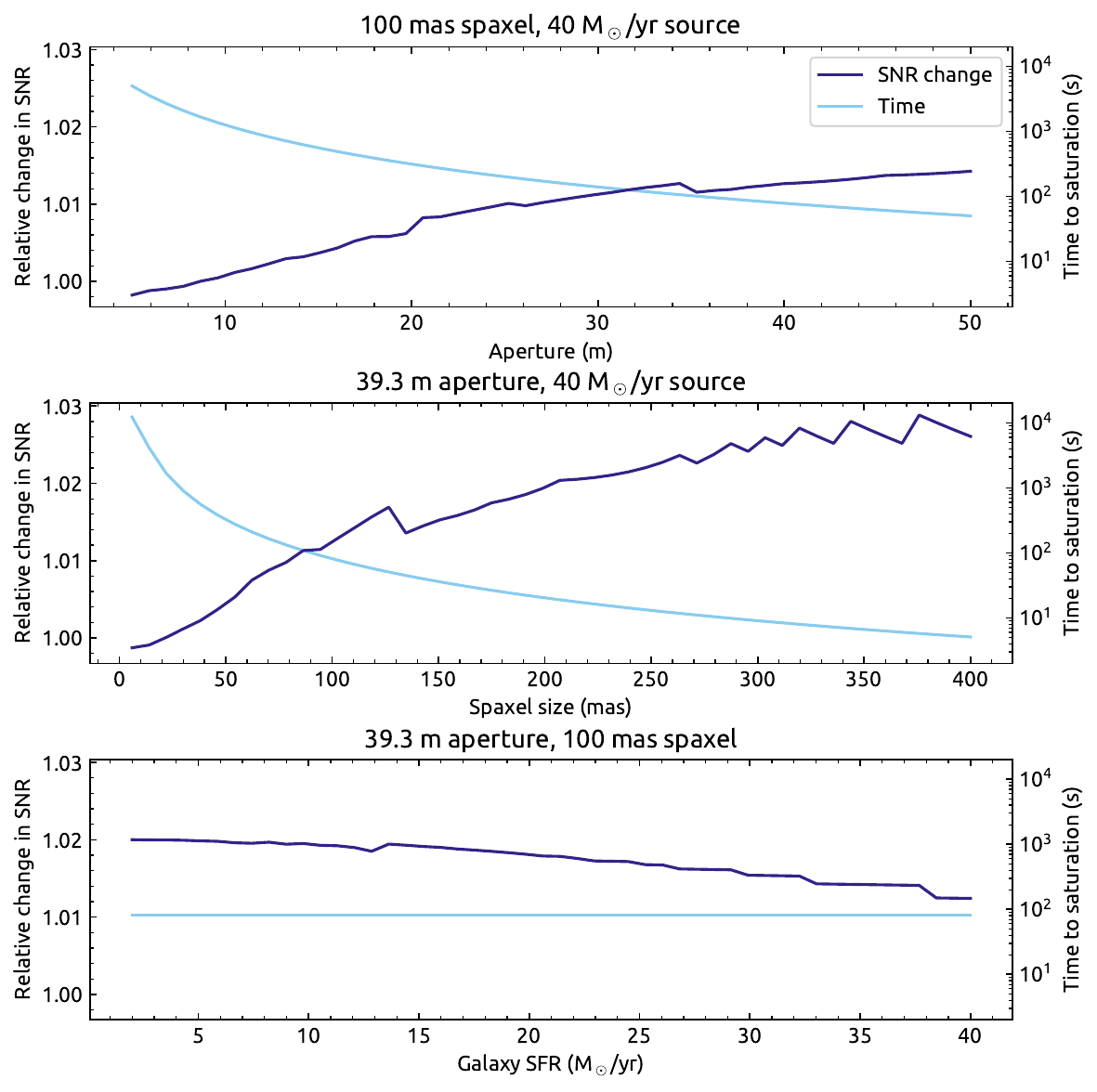}
    \caption{Simulated improvements in SNR due to using BRR relative to conventional observations as a function of telescope aperture, side length of a square spaxel, and target brightness. Also shown is the time it takes for the brightest OH lines to saturate for each combination of aperture and spaxel size. The gains from BRR are greatest when time to saturation is short.}
    \label{fig:snr-changes}
\end{figure}

Based on Equation \ref{eq:snr}, using BRR to increase exposure times could increase SNR for faint targets between bright OH lines, such as the example above.  To quantify the impact, we individually vary the telescope aperture, spaxel size, and galaxy brightness for the mock spectra and calculate the number of exposures of maximum duration before saturation to reach a peak SNR of 5. Since BRR prevents saturation, we then calculate how applying BRR would impact the SNR by simulating the same total exposure time using as few individual exposures as possible, up to a single exposure time of 30~minutes. For each simulated observation, we assume the same throughput and noise properties as GIRMOS, as well as a read noise of 10~e$^-$, which is within the typical range of values for CDS on HxRG detectors, though read noise for UTR sampling can be lower still\cite{Blank2011}. Figure \ref{fig:snr-changes} shows the fractional change in SNR by using BRR relative to the conventional observing mode by varying individual parameters, along with the time to saturation for those combinations of parameters.

In our simulations, individually increasing telescope aperture and spaxel size and decreasing target brightness each yield small SNR improvements. Varying each parameter simultaneously, BRR can increase SNR by around 10\% in the case of a 40~m aperture and 400~mas spaxel observing H$\alpha$ from a 2~M$_\odot$/yr galaxy. The gains for existing and upcoming instruments are more modest, as premier facilities tend to have lower detector read noise and smaller spaxel sizes to take advantage of their adaptive optics systems. The BRR method may provide the greatest increase in SNR in cases where large apertures and spaxels lead to fast saturation of the OH lines, resulting in the need for large numbers of short exposures. This will have the most impact when read noise is high, which could result from instruments using more pixels per spectral resolution element, for example.

For instruments such as GIRMOS and HARMONI, the benefit of BRR lies in the increased ability to quickly and accurately measure OH lines and their variability for precise sky subtraction. BRR allows monitoring of the change in flux of an OH line over the course of an exposure, which in turn provides information about the simultaneous behaviour of other OH lines\cite{Ramsay1992,Davies2007}. Because of the high, fast variability of these lines, they may change significantly in flux during the exposure, which will degrade the sky subtraction, leaving residuals of a few percent of the sky signal (i.e., 100s of e$^-$).\cite{Davies2007,Sharp2010,Soto2016} It is therefore beneficial to monitor the behaviour of the lines by periodically fitting a new ramp to the UTR samples of an OH line. While this can be done with any UTR sampling scheme, BRR allows this to happen at a fast cadence for exposure lengths which may be longer than the saturation time of the lines. In addition, Equation \ref{eq:var} shows that simply measuring a bright OH line at an increased rate by placing a window on it during an exposure decreases the uncertainty associated with that measurement, which can improve the sky subtraction residuals. We therefore consider BRR to be a promising tool for enabling accurate subtraction of variable sky backgrounds by providing new information not normally available on the variation of sky lines during a scientific exposure. The incorporation of this new information into specialized sky subtraction algorithms needs further investigation outside the immediate scope of this current work.

\section{Summary and Conclusions}
\label{sec:conclusions}

In this paper, we have described and demonstrated an on-detector method for preventing saturation of bright sky lines in NIR spectroscopy which has the potential to improve sky subtraction. Existing strategies to compensate for these bright, variable lines involve frequent, offset sky exposures or specialized optics that remove all information at the wavelengths of the emission lines. The BRR method described here is capable of tracking the changing flux of selected lines during an exposure while also preventing those lines from saturating the detector during long exposures. This is done by resetting the pixels of the detector containing the lines before they are able to saturate, and fitting a new flux to those pixels each time they are reset.

Our demonstration shows that we do not expect the BRR mode to produce any adverse systematic effects on observations in a well-controlled environment. The increase in flux caused by BRR in this test is explained by a lack of nonlinearity correction, which is due to the limited umber of UTR samples in our experiment. This correction would be straightforward to implement on dedicated facilities. As a result, we consider our experiment to be a successful test of this technique which shows potential for improved observations at little cost.

The BRR mode could be useful for improving SNR in read noise--limited observations by allowing unsaturated long exposures instead of stacking many short exposures. However, we estimate that, for premier facilities, this improvement is relatively small, and instruments with higher read noise and larger spaxel sizes will see greater gains. For most instruments, BRR offers an advantage by measuring time-resolved information on the varying sky background which is not typically available during the science exposure, which we expect will be beneficial to the development of more precise sky subtraction algorithms. BRR is also capable of increasing SNR for sky lines which are being read by windows as a result of the large number of samples for those pixels in each exposure. These improvements come without the need for any additional specialized hardware, as BRR is implemented completely in software/firmware.

Further development and implementation on an existing spectrograph will allow us to demonstrate the performance of this scheme on on-sky observations. Doing so will directly show the advantages that we expect to achieve by using this technique. We believe that BRR has the potential to make significant improvements for next-generation spectrographs, enabling key science on the faintest objects observable.

\subsection*{Disclosures}
The authors have no financial or other potential conflicts of interest to disclose.

\subsection*{Code and Data Availability} 
Code used for analysis and select data products are publicly available at \url{https://github.com/tgrosson/public-code/tree/main/Bright%20Region%20Reset}. The raw data are stored in a private directory on the Canadian Advanced Network for Astronomical Research (CANFAR, \url{https://www.canfar.net/}) and can be made available on reasonable request to the corresponding author.

\subsection*{Acknowledgments}

We thank the reviewers for their comments, which substantially improved the manuscript. This work is an expansion of our 2024 Proceedings paper\cite{Grosson2024}. We would like to thank Suresh Sivanandam and Adam Muzzin for their helpful discussions. We acknowledge the support of the Natural Sciences and Engineering Research Council of Canada (NSERC) Discovery Grants program. TAG acknowledges support from the NSERC-CREATE New Technologies for Canadian Observatories (NTCO) program.


\bibliography{main}   
\bibliographystyle{spiejour}   


\vspace{2ex}\noindent\textbf{Theodore A. Grosson} is a PhD student in astronomy at the University of Victoria in British Columbia, Canada. Their research focuses on instrumentation for astronomy, especially detector technologies and new techniques for improving observations. Scientifically, they are interested in the properties of dwarf galaxies outside the Local Group.

\vspace{1ex}
\noindent Biographies of the other authors are not available.

\end{document}